\documentclass[twocolumn]{aastex63}

\usepackage{amsmath}
\usepackage{array}
\usepackage{enumitem}
\newcolumntype{C}[1]{>{\centering\arraybackslash}p{#1}}
\newcolumntype{L}[1]{>{\raggedright\let\newline\\\arraybackslash\hspace{0pt}}m{#1}}

\received{July 8, 2021}
\revised{July 14, 2021}
\accepted{July 14, 2021}

\shorttitle{Phase Separation in Ultramassive White Dwarfs}
\shortauthors{Blouin \& Daligault}

\begin{document}

\title{Phase Separation in Ultramassive White Dwarfs}

\correspondingauthor{Simon Blouin}
\email{sblouin@lanl.gov}

\author[0000-0002-9632-1436]{Simon Blouin}
\affiliation{Los Alamos National Laboratory, PO Box 1663, Los Alamos, NM 87545, USA}

\author[0000-0002-8844-6124]{J{\'e}r{\^o}me Daligault}
\affiliation{Los Alamos National Laboratory, PO Box 1663, Los Alamos, NM 87545, USA}

\begin{abstract}
Ultramassive white dwarfs are extreme endpoints of stellar evolution. Recent findings, such as a missing multi-Gyr cooling delay for a number of ultramassive white dwarfs and a white dwarf with a quasi-Chandrasekhar mass, motivate a better understanding of their evolution. A key process still subject to important uncertainties is the crystallization of their dense cores, which are generally assumed to be constituted of $^{16}$O, $^{20}$Ne, and a mixture of several trace elements (most notably $^{23}$Na and $^{24}$Mg). In this work, we use our recently developed Clapeyron integration technique to compute accurate phase diagrams of three-component mixtures relevant to the modeling of O/Ne ultramassive white dwarfs. We show that, unlike the phase separation of $^{22}$Ne impurities in C/O cores, the phase separation of $^{23}$Na impurities in O/Ne white dwarfs cannot lead to the enrichment of their cores in $^{23}$Na via a distillation process. This severely limits the prospect of transporting large quantities of $^{23}$Na toward the center of the star, as needed in the white dwarf core collapse mechanism recently proposed by Caiazzo et al. We also show that despite representing $\approx 10\%$ of the ionic mixture, $^{23}$Na and $^{24}$Mg impurities only have a negligible impact on the O/Ne phase diagram, and the two-component O/Ne phase diagram can be safely used in white dwarf evolution codes. We provide analytic fits to our high-accuracy O/Ne phase diagram for implementation in white dwarf models.
\end{abstract}

\section{Introduction}
Ultramassive white dwarfs are generally defined as white dwarfs with masses $\gtrsim 1.1\,M_{\odot}$ \citep{vennes2008,camisassa2019}. This is significantly above the prominent peak of the white dwarf mass distribution at $\approx 0.6\,M_{\odot}$ \citep{bergeron2019,kepler2019}, meaning that those objects are quite rare. They represent 2.3\% of the 100 pc white dwarf sample in the Sloan Digital Sky Survey footprint \citep{kilic2020}, and not a single one is known within 20~pc of the Sun \citep{hollands2018}. But being the prospective progenitors of type Ia supernovae and the potential remnants of double white dwarf mergers, reasons abound to study those objects. Thanks to the Gaia mission \citep{gaiadr2a,gaiadr2b,gentile2019}, the number of known ultramassive white dwarfs is increasing rapidly \citep{kilic2021}. This has allowed the discovery of a number of peculiar ultramassive white dwarfs \citep{hollands2020,pshirkov2020,caiazzo2021} and the compilation of a large enough sample to study their evolution in unprecedented detail \citep{cheng2019}. 

Ultramassive white dwarfs can be formed through the single-star evolution of a $\approx 8-10\,M_{\odot}$ star \citep{siess2007,camisassa2019} or following the merger of two normal-mass white dwarfs \citep{loren2009,dan2014,cheng2020,temmink2020}. In both cases, current evolution models predict that the resulting white dwarf should have a core mostly made of $^{16}$O and $^{20}$Ne \citep{siess2007,siess2010,schwab2021}. At odds with those predictions, observational evidence is mounting that at least some ultramassive white dwarfs have C/O cores instead \citep{bauer2020,blouin2021a,camisassa2021}. \cite{cheng2019} identified a population of ultramassive white dwarfs that undergo an additional cooling delay of $\sim 8\,$Gyr compared to the predictions of standard cooling models. This delayed population is located on the so-called Q branch of the Gaia color--magnitude diagram \citep{gaiaQbranch}. This branch corresponds to the predicted location of the crystallization of C/O cores \citep{tremblay2019}, which is distinct from the predicted location of O/Ne crystallization (which occurs at higher temperatures given the higher ionic charges of the plasma). Some evolutionary scenarios have been proposed to explain the existence of ultramassive C/O white dwarfs \citep{althaus2021}, but the relative sizes of the C/O and O/Ne ultramassive white dwarf populations remain an open question.

We recently showed that the phase separation (or fractionation) of trace amounts of $^{22}$Ne during the crystallization of ultramassive C/O white dwarfs can provide the required energy source to explain \citeauthor{cheng2019}'s missing cooling delay \citep{blouin2021a}. During the solidification of a multicomponent plasma, the coexisting solid and liquid phases generally have different compositions. This composition change is given by the phase diagram of the ionic mixture, which depends on the charges of the ions. For compositions relevant to C/O white dwarfs, we found that the $^{22}$Ne abundance is lower in the solid than in the liquid phase. Due to the extra neutrons of the $^{22}$Ne isotope ($A > 2Z$), this $^{22}$Ne deficit can render the solid crystals lighter than the surrounding liquid. The solid crystals that are formed near the center of the white dwarf (where the conditions for solidification are first met) are therefore expected to float and melt in lower density regions above the central layers \citep{isern1991}. The constituent ions of those $^{22}$Ne-poor crystals are then mixed in the liquid layers, and $^{22}$Ne-rich liquid is gradually displaced toward the center of the white dwarf. This distillation process is a very efficient way of liberating the gravitational energy stored in $^{22}$Ne, which can naturally explain the missing cooling delay.

Ultramassive white dwarfs with O/Ne cores are predicted to contain a sizeable amount of $^{23}$Na, with an abundance mass fraction of $X(^{23}{\rm Na}) \approx 0.05-0.06$ \citep{camisassa2019,schwab2021b}. Due to the neutron-rich nature of $^{23}$Na and the similarity between the charge ratios of C/O/Ne and O/Ne/Na mixtures, a $^{23}$Na distillation process similar to that described above for $^{22}$Ne could a priori take place in O/Ne white dwarfs. Not only could this represent a missing cooling delay in current evolutionary models of ultramassive white dwarfs, but it could also be the catalyst of the new white dwarf collapse mechanism proposed by \cite{caiazzo2021}. 

At the extreme densities that characterize the central layers of the most massive white dwarfs, \cite{caiazzo2021} point out that the nuclei of some elements can undergo electron capture \citep[see also][]{salpeter1961,shapiro1983}. This process would remove electrons from the plasma, thereby reducing the degeneracy pressure that supports the star. In the case of the extreme white dwarf ZTF~J190132.9+145808.7, \cite{caiazzo2021} explain that if at least 60\% of the $^{23}$Na sinks deep enough to undergo electron capture, the star could collapse and form either a neutron star or a supernova. Due to its additional neutron, $^{23}$Na is expected to gradually sink toward the center of the white dwarf \citep{bildsten2001}. However, this transport process is halted in the solidified layers of the core \citep{hughto2011}. \cite{schwab2021b} showed that since ultramassive O/Ne white dwarfs crystallize very early in their evolution (due to the extreme density of their cores), it is highly unlikely that a significant amount of $^{23}$Na can be transported to the central layers by this gravitational settling process \citep[see also][]{camisassa2021}. On the other hand, the distillation of $^{23}$Na, if it takes place, could be a very efficient way of transporting $^{23}$Na to the central layers that could make \citeauthor{caiazzo2021}'s electron capture collapse mechanism possible. Contrary to gravitational settling, $^{23}$Na distillation would not be hampered by crystallization, but instead triggered by it.

In this work, we use our state-of-the-art Clapeyron phase diagram calculation technique to find out whether $^{23}$Na distillation can take place in ultramassive O/Ne white dwarfs. We also investigate the extent to which the two-component O/Ne phase diagram currently used in white dwarf evolution models is a good approximation of the phase diagram of the real multicomponent mixture, which includes non-negligible traces of $^{23}$Na and $^{24}$Mg. In Section~\ref{sec:ONe}, we present our calculation of the two-component O/Ne phase diagram. This calculation is a necessary intermediate step for our calculation of the three-component O/Ne/Na and O/Ne/Mg phase diagrams. We also provide analytic fits to this phase diagram to facilitate its implementation in white dwarf models. In Section~\ref{sec:3cp}, we then present our three-component O/Ne/Na and O/Ne/Mg phase diagrams, which we use to investigate the possibility of $^{23}$Na distillation and to assess the impact of impurities on the two-component O/Ne phase diagram. Finally, our conclusions are given in Section~\ref{sec:conclu}.

\section{The Two-Component Oxygen/Neon phase diagram}
\label{sec:ONe}
\subsection{Methods}
Our approach to calculate the O/Ne phase diagram is identical to that used in \cite{blouin2020} and described at length in \cite{blouin2021b}. It is based on the Clapeyron (or Gibbs--Duhem) method \citep{kofke1993a,kofke1993b,hitchcock1999}. The Clapeyron equations are differential equations that describe the phase boundaries in the space of intensive thermodynamic variables. The idea of the Clapeyron method is to simply integrate the relevant Clapeyron equation to obtain the phase boundary between two coexisting phases. For a two-component phase diagram, we perform this integration at constant pressure in the temperature--chemical potential difference space [\citealt{blouin2021b}, Equation (10)]. This integration requires the calculation of enthalpies and concentrations for fixed pressures, temperatures, and chemical potential differences between both ionic species. We obtain those quantities using Monte Carlo simulations in the isobaric semi-grand canonical ensemble (NPT$\Delta \mu$). We refer the reader to \cite{blouin2021b} for the theoretical and numerical details of this advanced technique.

Throughout this work, all our Monte Carlo simulations include $N=686$ ions and are executed for $7 \times 10^6$ iterations, the first $2 \times 10^6$ being discarded from the calculations of the averages to avoid the initial equilibration phase. The interaction between ions is assumed to take the form of a Yukawa (screened Coulomb) potential, with a screening length given by the long-wavelength approximation and adjusted according to the electron density in the plasma \citep{blouin2021b}. The electron background is also explicitly included in our simulations, which is necessary due to the constant-pressure nature of our approach. This is to be contrasted with more standard techniques \citep{horowitz2010,medin2010}, where a constant-volume approximation is usually made. We assume a fixed pressure of $10^{24}\,{\rm erg\,cm}^{-3}$, which is representative of white dwarf cores \citep{fontaine2001}. We have previously investigated the sensitivity of the C/O phase diagram to the pressure and found no important effect within the range of pressures relevant to white dwarf interiors \citep{blouin2021b}. This is consistent with independent results \citep{medin2010} and stems from the fact that volume changes upon solidification are very small in dense plasmas, where the degenerate electron background completely dominates the pressure

\subsection{Results}
\begin{figure}
    \includegraphics[width=\columnwidth]{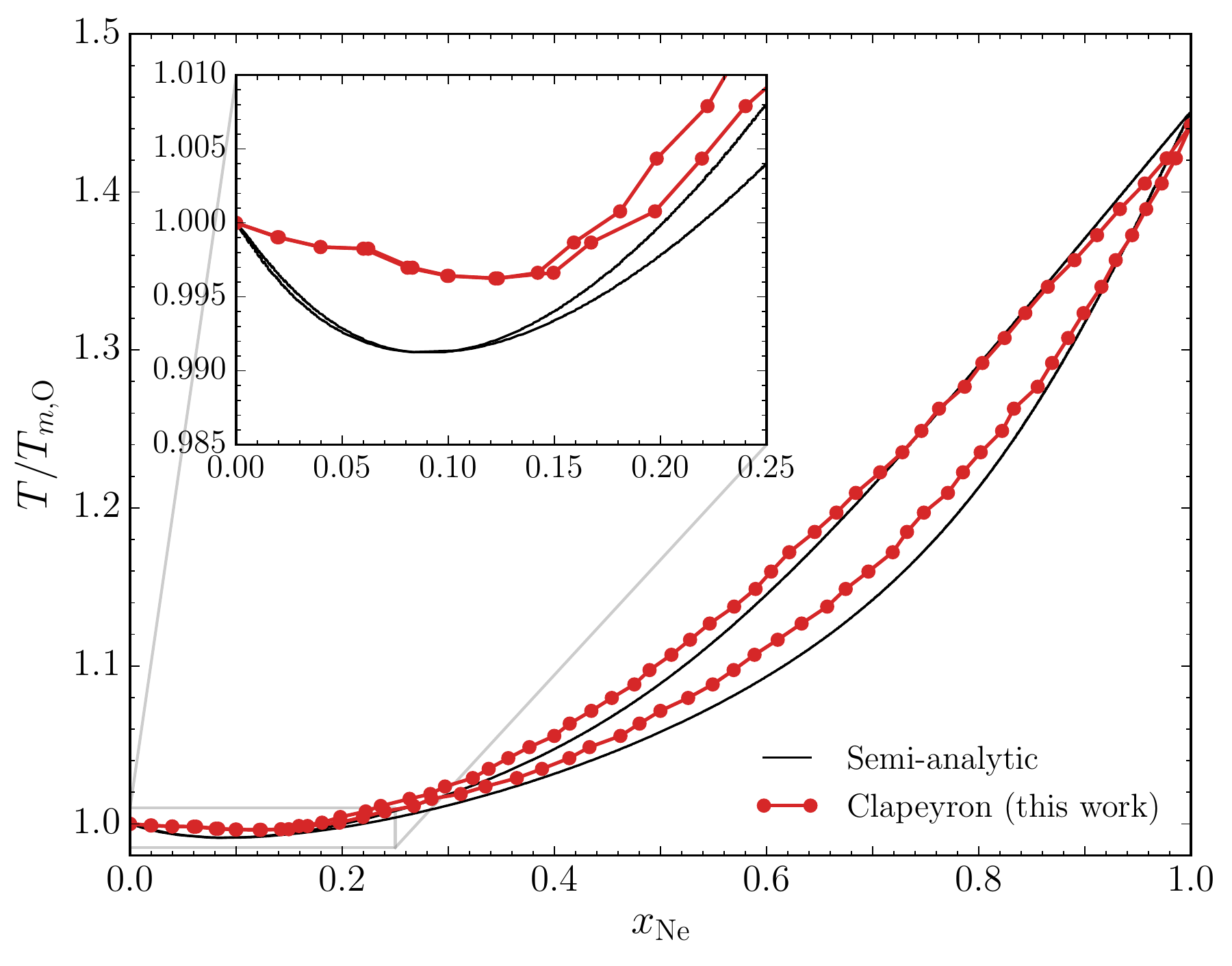}
    \caption{O/Ne phase diagram. The upper curve is the liquidus (above which the O/Ne mixture is always liquid), and the lower curve is the solidus (below which the plasma forms a bcc solid). The vertical axis gives the temperature in units of the melting temperature of a pure O plasma ($\Gamma=178$), and the horizontal axis gives the Ne number concentration of the mixture. The inset zooms in on the region where an azeotrope is predicted. Our results (shown in red) are compared to those obtained using the semi-analytic approach described in \cite{medin2010}.}
  \label{fig:ONe}
\end{figure}

Figure~\ref{fig:ONe} shows our O/Ne phase diagram. The charge ratio of an O/Ne mixture being close to that of a C/O mixture ($Z_2 / Z_1= 1.25$ and 1.33, respectively), it is unsurprising to see an azeotrope shape similar to that of the C/O phase diagram. As described in \cite{camisassa2019}, the shape of the O/Ne phase diagram implies a phase separation process that enriches in Ne the central layers of crystallizing ultramassive white dwarfs. For comparison, we also show in Figure~\ref{fig:ONe} the O/Ne phase diagram predicted by the semi-analytic approach of \cite{medin2010}.\footnote{We use the code provided at \url{https://github.com/andrewcumming/phase_diagram_3CP} to calculate the semi-analytic phase diagram.} This second phase diagram is in very good agreement with our calculations, just as for the C/O case \citep{blouin2020}. Note that the small disagreements apparent in Figure~\ref{fig:ONe} are to be expected given a number of approximations in the \cite{medin2010} approach that we do not make with the Clapeyron technique (e.g., no screening of the ion--ion interactions, use of a pure linear mixing rule for the calculation of the free energy of the liquid phase, implicit assumption that the vibration modes of the two-component zero-temperature solid are identical to that of the one-component zero-temperature solid). 

\subsection{Analytic fits}
\begin{figure*}
    \centering
    \includegraphics[width=\columnwidth]{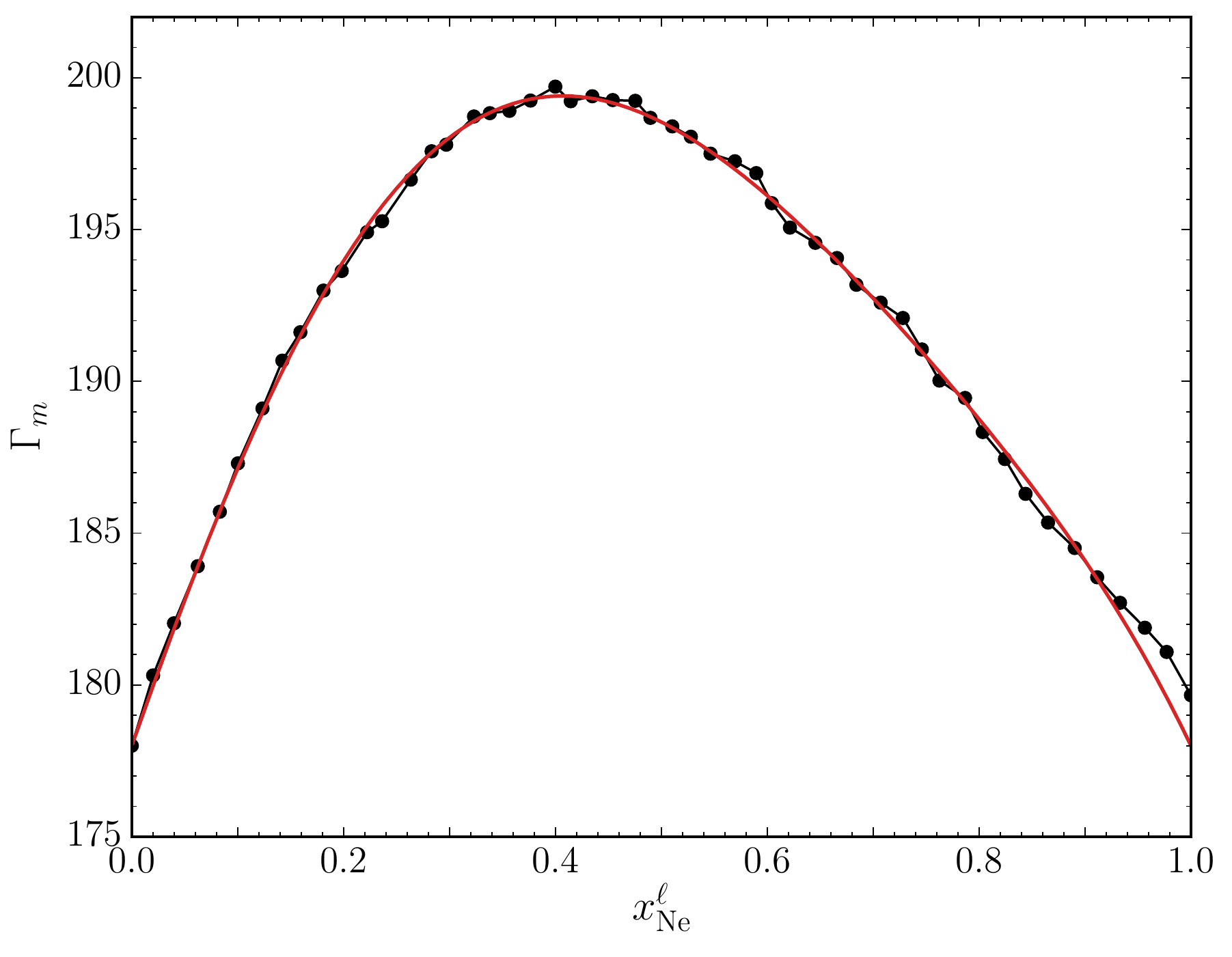}
    \includegraphics[width=\columnwidth]{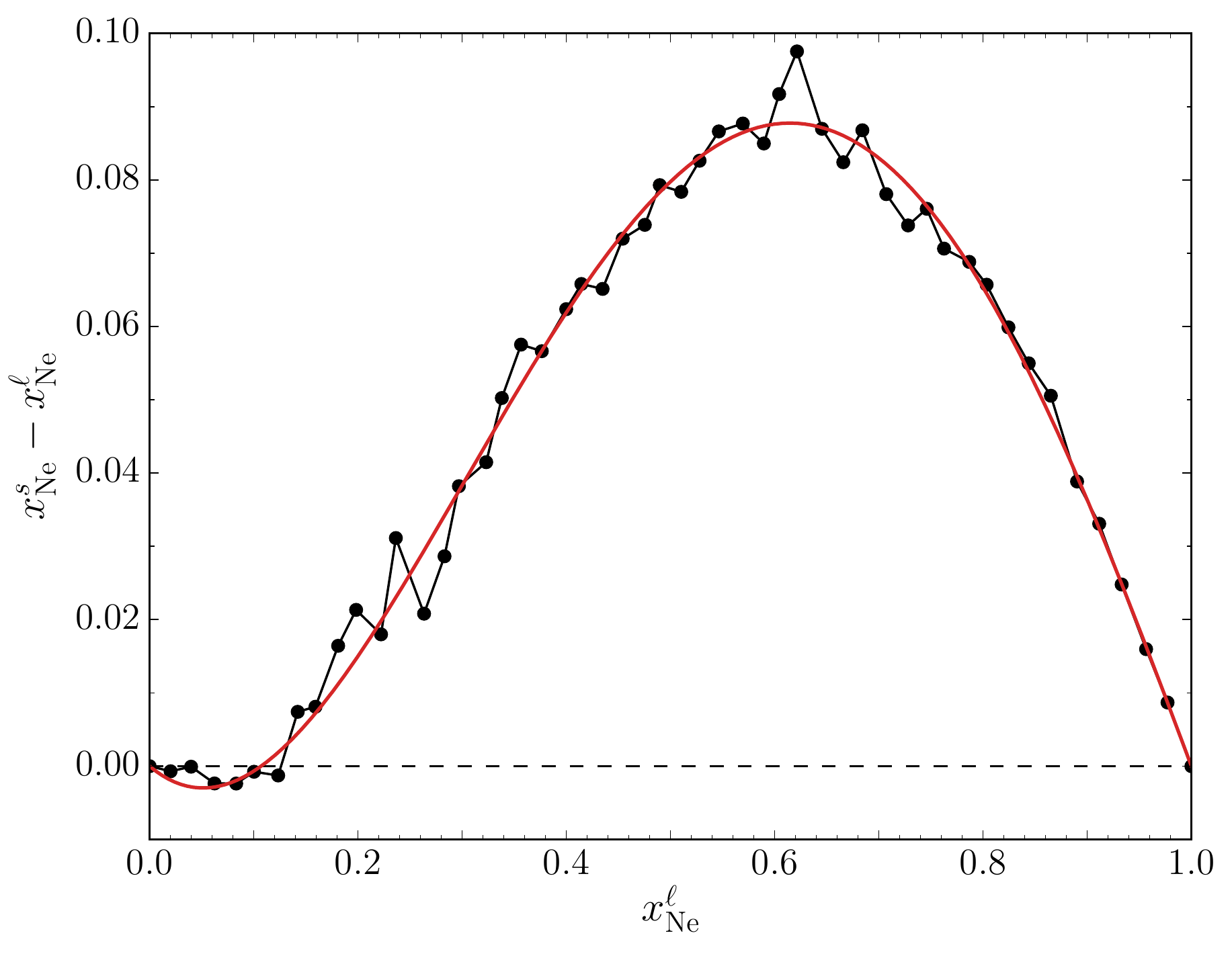}
    \caption{{\it Left}: Coupling parameter [as defined by Equation~(\ref{eq:gamma})] at which an O/Ne mixture with Ne number concentration $x_{\rm Ne}^{\ell}$ crystallizes. {\it Right}: Difference in Ne concentration between the coexisting solid and liquid phases as a function of the Ne concentration in the liquid phase. For both panels, the data in black was extracted from our results of Figure~\ref{fig:ONe}, and the red lines correspond to analytic fits [Equation~\eqref{eq:fit}].}
  \label{fig:ONe_fit}
\end{figure*}

We regard our O/Ne phase diagram as the most accurate version of this calculation published to date. This phase diagram is important in ultramassive white dwarf models to account for the O/Ne phase separation \citep{isern1997,geronimo2018,camisassa2019}. We therefore encourage its implementation in white dwarf codes, which we hope to facilitate by providing analytic fits to the coupling parameter of the mixture at the phase transition, $\Gamma_m$, and to the composition change, $\Delta x_{\rm Ne} = x_{\rm Ne}^s - x_{\rm Ne}^{\ell}$, as a function of the Ne number concentration in the liquid phase, $x_{\rm Ne}^{\ell}$. We use the standard definition for the coupling parameter of the mixture,
\begin{equation}
\Gamma = \frac{ \langle Z^{5/3} \rangle e^2}{a_e k_B T},
\label{eq:gamma}
\end{equation}
where $\langle Z^{\alpha} \rangle = \sum_i Z_i^{\alpha} n_i / \sum_i n_i$ (the sums run over all ionic species) and $a_e = (3/4 \pi n_e)^{1/3}$, with $Z_i e$ the charge of ionic species $i$, $n_i$ the number density of $i$, $n_e = \sum_i Z_i n_i$ the electron density, $k_B$ the Boltzmann constant, and $T$ the temperature. As with the C/O phase diagram \citep{blouin2021b}, we find that a fifth-order polynomial,
\begin{equation}
\sum_{i=0}^5 a_i (x_{\rm Ne}^{\ell})^i,
\label{eq:fit}
\end{equation}
can satisfactorily reproduce $\Gamma_m (x_{\rm Ne}^{\ell})$ and $\Delta x_{\rm Ne} (x_{\rm Ne}^{\ell})$. The fit coefficients $a_i$ are given in Table~\ref{tab:fit}. As shown in Figure~\ref{fig:ONe_fit}, the analytic fits reproduce our simulation data within their statistical noise. Note that the fits were forced to reproduce the known one-component limits $\Gamma_m = 178$ and $\Delta x_{\rm Ne}=0$ at $x_{\rm Ne}^{\ell}=0$ and 1.

\begin{table}
\begin{ruledtabular}
\begin{tabular}{lrr}
 & $\Gamma_m (x_{\rm Ne}^{\ell})$ & $\Delta x_{\rm Ne} (x_{\rm Ne}^{\ell})$\\
\hline
$a_0$ & 178.000000 & 0.000000 \\ 
$a_1$ & 99.175544 & $-$0.120299\\ 
$a_2$ & $-$53.498901 & 1.304399\\ 
$a_3$ & $-$292.291988 & $-$1.722625\\ 
$a_4$ & 421.150375 & 0.393996 \\ 
$a_5$ & $-$174.534905 & 0.144529 \\ 
\end{tabular}
\end{ruledtabular}
\caption{\label{tab:fit} Fit parameters for $\Gamma_m (x_{\rm Ne}^{\ell})$ and $\Delta x_{\rm Ne} (x_{\rm Ne}^{\ell})$ [Equation~\eqref{eq:fit}].}
\end{table}

For consistency with the fit previously given for the C/O phase diagram \citep{blouin2021b}, we fitted the crystallization temperature in term of the coupling parameter $\Gamma$ of the mixture. Alternatively, the crystallization temperature can be expressed as the coupling parameter of one ionic component in the mixture, e.g.,
\begin{equation}
\Gamma_{\rm O} = \frac{ Z_{\rm O}^{5/3} e^2}{a_e k_B T}.
\end{equation}
As we will show in Section~\ref{sec:impurities}, the $\Gamma_{{\rm O},m}$ of a two-component O/Ne mixture is very close to that of an O/Ne mixture with impurities. Because $\langle Z^{5/3} \rangle$ is different in the cases with and without impurities,\footnote{Note that $a_e$ remains virtually unchanged because a constant pressure implies a nearly constant electron density in the strongly degenerate limit.} our fit to $\Gamma_m$ as defined by Equation~\eqref{eq:gamma} should not be used to calculate the crystallization temperature of an O/Ne mixture when impurities are also included. In the latter case, our analytic fit to the $\Gamma_m$ of the two-component mixture should first be converted into $\Gamma_{{\rm O},m}$,
\begin{equation}
\Gamma_{{\rm O},m} = \frac{\Gamma_m Z_{\rm O}^{5/3}}{x_{\rm O} Z_{\rm O}^{5/3} + x_{\rm Ne} Z_{\rm Ne}^{5/3}},
\end{equation}
and then compared to the actual $\Gamma_{\rm O}$ of the mixture to assess whether the plasma is solid or not. The same considerations apply to our fit of the C/O phase diagram \citep{blouin2021b}.

\section{Three-Component Phase Diagrams}
\label{sec:3cp}
\subsection{Methods and validation}
To calculate three-component phase diagrams, we use the same approach as in \cite{blouin2021a}. The integration is performed at constant pressure and temperature in the space spanned by the chemical potential differences between the three ionic species [Equation~(A2), \citealt{blouin2021b}]. The initial conditions at zero impurity concentration are given by the O/Ne phase diagram presented above (see Appendix~A of \citealt{blouin2021b} for details). To fully map a three-component phase diagram at a fixed pressure in the three-dimensional temperature--composition space, we need to perform many distinct constant-temperature Clapeyron integrations.

\begin{figure}
    \includegraphics[width=\columnwidth]{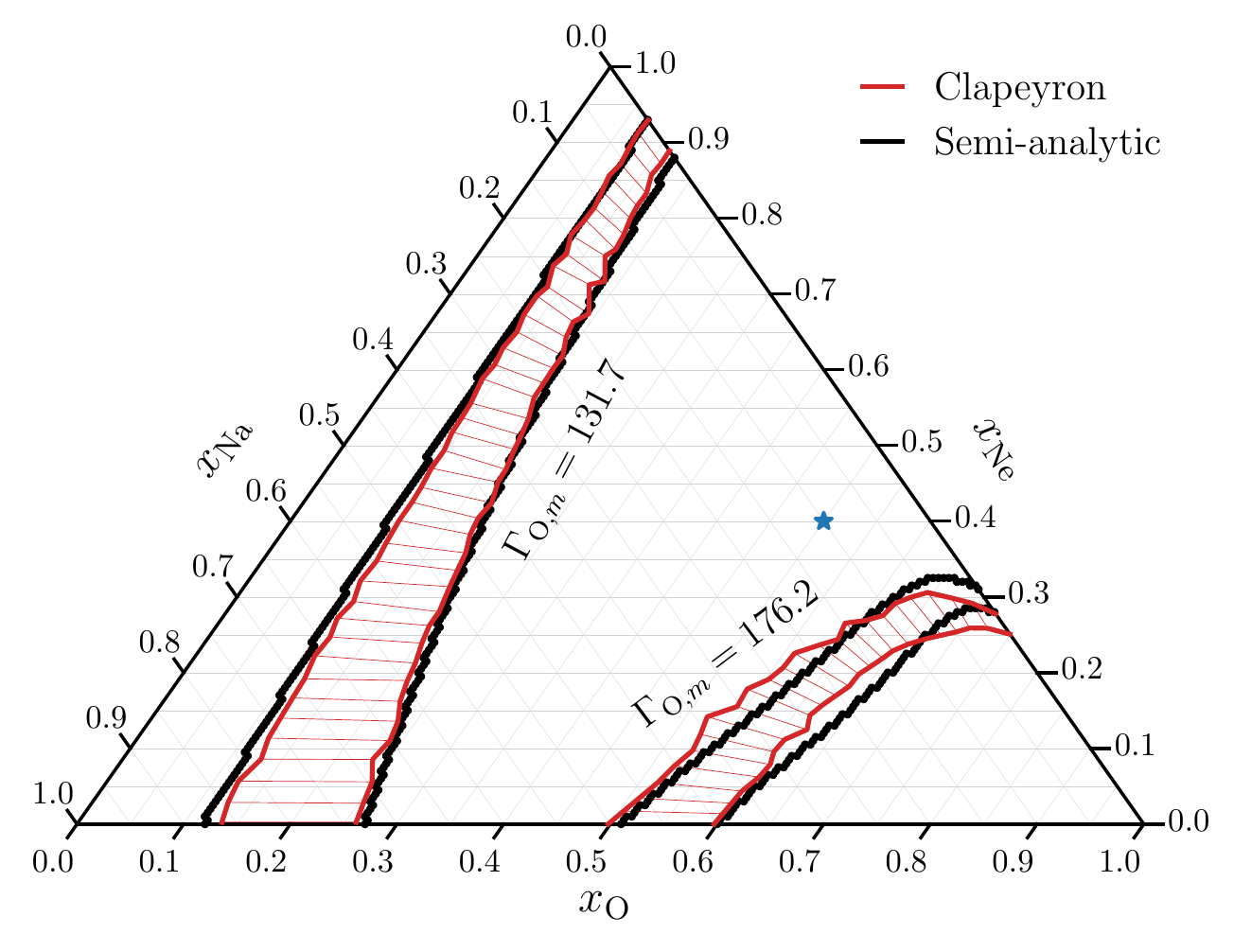}
    \caption{O/Ne/Na phase diagram. The complete constant-pressure three-component phase diagram lives in the $x_{\rm O}$-$x_{\rm Ne}$-$x_{\rm Na}$-$T$ space. Here, we project it in the $x_{\rm O}$-$x_{\rm Ne}$-$x_{\rm Na}$ space for two fixed $T$'s (identified in the figure in terms of $\Gamma_{\rm O}$). For both temperatures, the upper curve is the solidus and the lower one is the liquidus. The thin red lines between the liquidus and the solidus connect the compositions of the coexisting liquid and solid phases (i.e., they trace the composition change upon solidification). The composition of a given mixture can be obtained by following the thin gray lines to the tick marks that have the same slopes. As an example, the blue star corresponds to $(x_{\rm O},x_{\rm Ne},x_{\rm Na})=(0.5,0.4,0.1)$. As in Figure~\ref{fig:ONe}, we compare our results to the semi-analytic approach of \cite{medin2010}.}
  \label{fig:ONeNa}
\end{figure}

Figure~\ref{fig:ONeNa} presents two constant-temperature integrations of the O/Ne/Na phase diagram. Integrating the coexistence lines up to very high Na concentrations is not directly relevant to white dwarfs (Na being only a trace species in O/Ne cores), but allows an instructive comparison between our method and the semi-analytic approach of \cite{medin2010} (see also \citealt{caplan2018,caplan2020}).\footnote{Again, we use the code provided at \url{https://github.com/andrewcumming/phase_diagram_3CP} to calculate the semi-analytic O/Ne/Na phase diagram.} Figure~\ref{fig:ONeNa} shows that the agreement between both approaches is excellent. The largest discrepancies appear for the lower temperature case ($\Gamma_{{\rm O},m}=176.2$) and stem mostly from differences in the two-component O/Ne phase diagram. The starting point of this integration (right axis of Figure~\ref{fig:ONeNa}) corresponds to a region of the O/Ne phase diagram ($x_{\rm Ne} \sim 0.25-0.30$) where there is a small offset between the two calculations (Figure~\ref{fig:ONe}).

\subsection{Investigating the possibility of $^{23}$Na distillation}
Figure~\ref{fig:ONeNa} does not support the existence of $^{23}$Na distillation in O/Ne white dwarfs. The thin red lines that trace the composition change due to crystallization (see caption) appear to never lead to a $^{23}$Na-depleted solid. If the solid is not depleted in the neutron-rich $^{23}$Na isotope, then it cannot be lighter than the surrounding liquid, and a distillation process analogous to that of $^{22}$Ne in C/O white dwarfs cannot take place. 

\begin{figure}
    \includegraphics[width=\columnwidth]{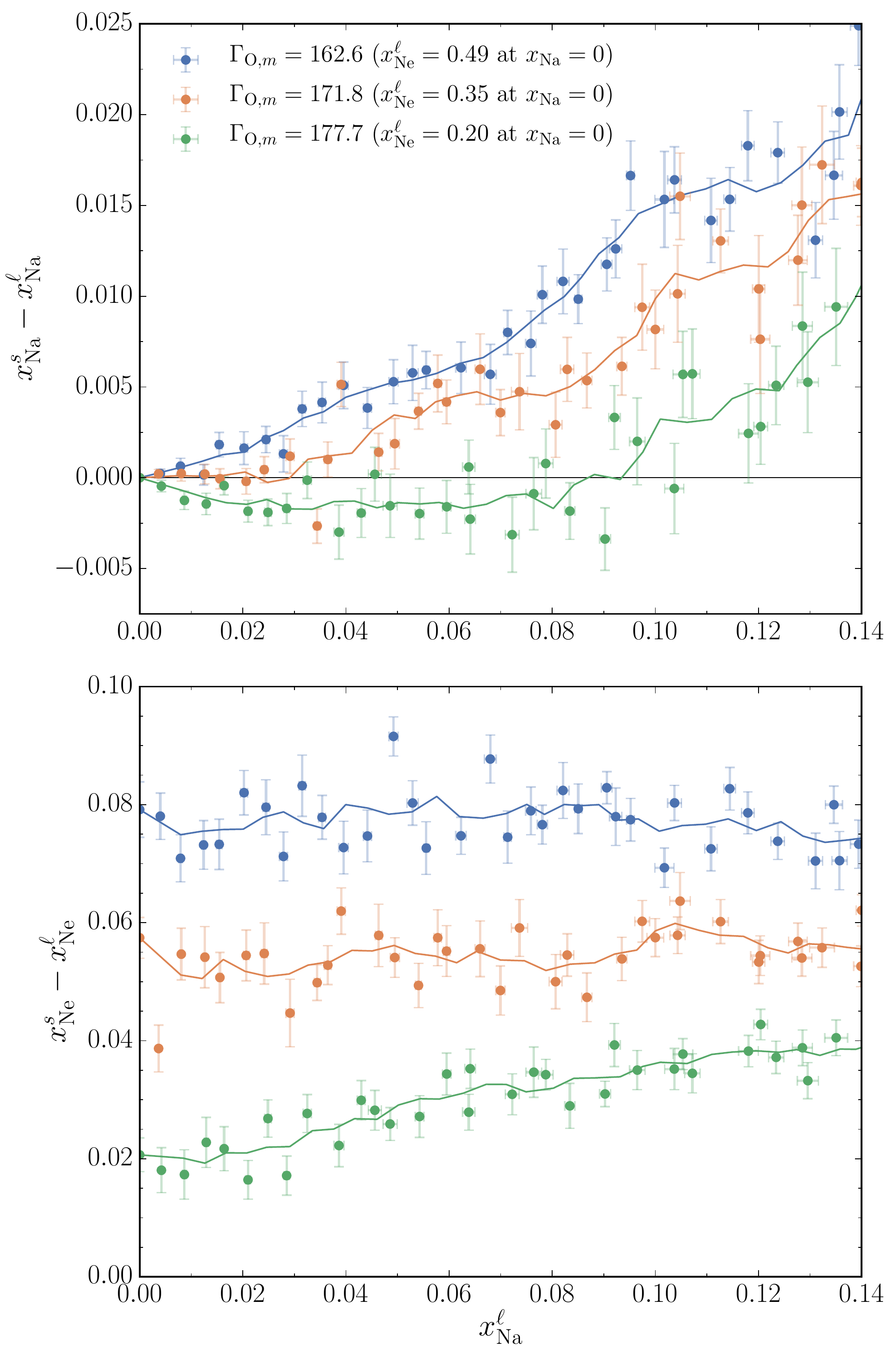}
    \caption{Na and Ne concentration changes at the liquid--solid phase transition as a function of the Na abundance in the liquid O/Ne/Na mixture. The results of three distinct integrations of the O/Ne/Na phase diagram at constant temperature and pressure are shown. Those different crystallization temperatures correspond to different Ne concentrations, as indicated in the legend. The error bars are the 1$\sigma$ confidence intervals obtained by applying the block-averaging technique to our Monte Carlo trajectories. The solid lines correspond to five-point moving averages.}
  \label{fig:deltax}
\end{figure}

\begin{figure}
    \includegraphics[width=\columnwidth]{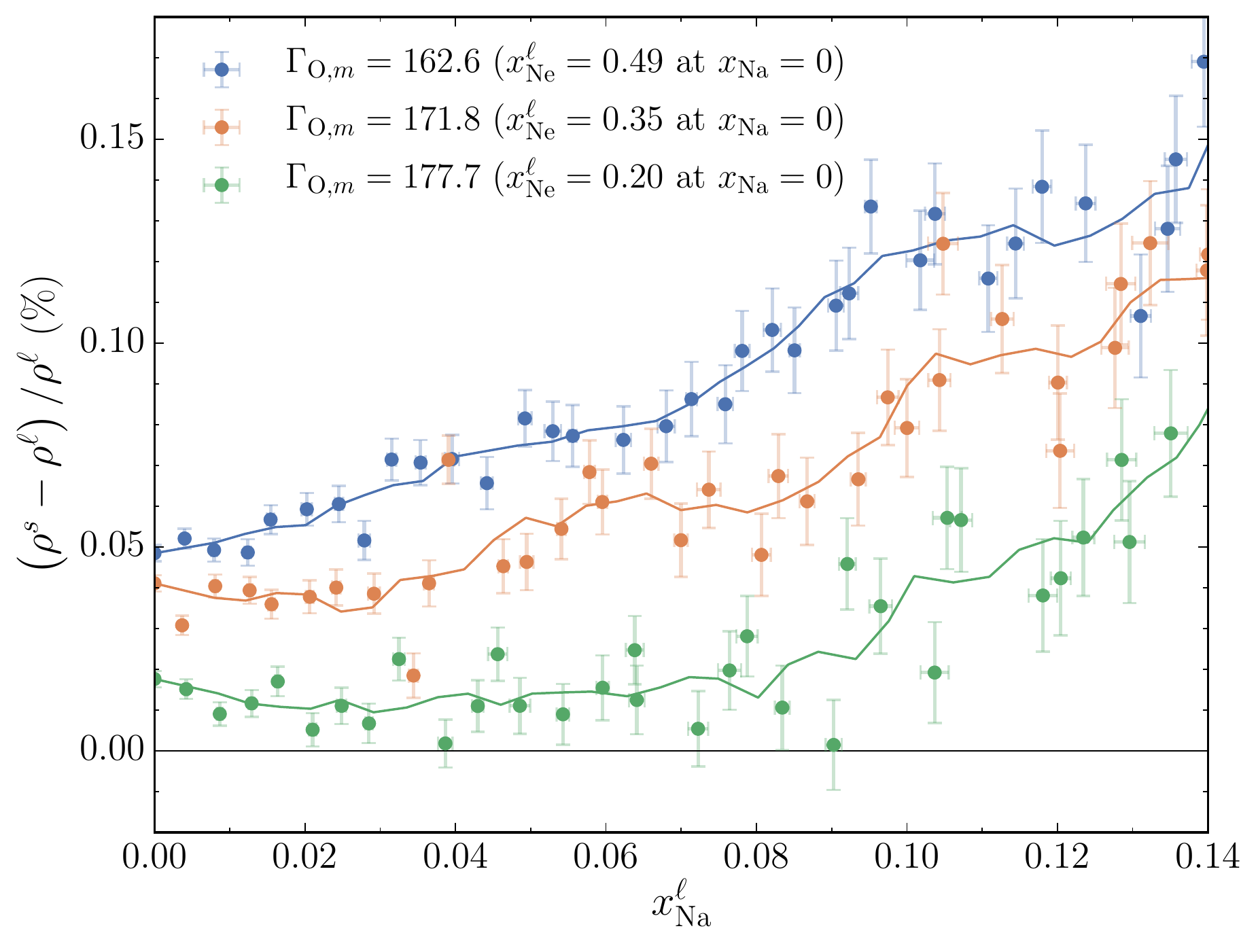}
    \caption{Relative mass density differences between the solid and liquid phases at the phase transition for the same constant-temperature integrations as in Figure~\ref{fig:deltax}.}
  \label{fig:rho_Na}
\end{figure}

To investigate this more closely, we performed eight additional constant-temperature integrations of the O/Ne/Na phase diagram, this time limiting ourselves to the low Na concentrations expected in O/Ne cores and using a finer integration grid than in Figure~\ref{fig:ONeNa}. We use temperatures ranging from $\Gamma_{{\rm O},m}=162.6$ to $178.6$; at $x_{\rm Na}=0$, this corresponds to Ne concentrations of $x_{\rm Ne}=0.49$ to $0.15$, respectively. This range includes and extends beyond the Ne concentrations expected in O/Ne white dwarfs. A typical O/Ne core composition is $\approx57$\%~$^{16}$O, 32\%~$^{20}$Ne, 6\%~$^{23}$Na, 3\%~$^{24}$Mg, and 2\% of other traces (most importantly $^{12}$C and $^{22}$Ne) by mass \citep{schwab2021b}, corresponding to $x_{\rm O}\approx 0.63$, $x_{\rm Ne}\approx 0.28$, $x_{\rm Na}\approx 0.05$, and $x_{\rm Mg}\approx 0.02$. Figure~\ref{fig:deltax} shows the composition change of the plasma upon crystallization as a function of the Na concentration in the liquid phase. Note that the O and Ne concentrations in the liquid phase are not constant throughout a given constant-temperature integration (see Figure~\ref{fig:ONeNa}). However, since Figure~\ref{fig:deltax} focuses on relatively small Na concentrations, the O and Ne concentrations at $x_{\rm Na}^{\ell}=0$ (given in the legend of Figure~\ref{fig:deltax}) are representative of the O and Ne concentrations across this limited range of $x_{\rm Na}^{\ell}$. The green line of Figure~\ref{fig:deltax} reveals that a very small $^{23}$Na depletion of the solid phase may be possible for O/Ne/Na plasmas with Ne concentrations somewhat lower than expected by current stellar evolution models. However, Figure~\ref{fig:rho_Na} shows that this slight $^{23}$Na depletion is insufficient to make the solid lighter than the liquid. The Ne enrichment of the solid is enough to counterbalance the effect of the $^{23}$Na depletion and to keep the solid heavier, preventing the existence of a distillation process. 

In our mass density ($\rho$) calculations, we have assumed that 3\% of the Ne is under its neutron-rich $^{22}$Ne isotopic form and that the rest is $^{20}$Ne, $X(^{22}{\rm Ne})/X(^{20}{\rm Ne})=0.03$, which is consistent with the predictions of stellar evolution models \citep{siess2007,schwab2021b}. If we assume instead that all Ne is under the $^{20}$Ne form, $\rho^s - \rho^{\ell}$ decreases because the Ne-rich solid is deprived from the mass density increase it received from $^{22}$Ne. It then becomes possible to have solid crystals that are slightly lighter than the coexisting liquid, but only at Ne concentrations significantly below the predicted values. Therefore, barring substantial changes to our understanding of the composition of O/Ne cores, we can safely conclude that $^{23}$Na distillation does not take place in ultramassive O/Ne white dwarfs. Instead, phase separation in crystallizing O/Ne white dwarfs leads to a simple sedimentation process \citep{mochkovitch1983,isern1997} where the central layers are enriched in Ne, as already described by current models \citep{camisassa2019}. Without $^{23}$Na distillation and given that ultramassive O/Ne white dwarfs crystallize too early to allow any significant gravitational settling of $^{23}$Na \citep{schwab2021b}, we conclude that there is no identified mechanism that can transport enough $^{23}$Na to the central layers of ultramassive white dwarfs to provoke an electron capture induced collapse \citep{caiazzo2021}. 

\subsection{Other isotopes}
Only neutron-rich isotopes can be expected to lead to a distillation process, since only them can sufficiently affect the mass density of the solid to change the sign of $\rho^s - \rho^{\ell}$ compared to the two-component O/Ne case. To a good approximation, the electron density of a dense plasma at a given pressure is almost independent of the nature of the ionic species in the mixture. This is the direct consequence of the fact that ions only have a negligible contribution to the pressure, which is dominated by the degenerate electron gas in the cores of white dwarfs. This property implies that species that have the same $A/Z=2$ ratio as the dominant $^{16}$O and $^{20}$Ne ions have almost no effect on the mass density at a given pressure and temperature, since they provide the same mass per electron. The phase separation of $A/Z \neq 2$ isotopes is needed to markedly alter $\rho^s - \rho^{\ell}$. This is why we have only focused on the neutron-rich $^{23}$Na even though other trace species with $A/Z=2$ are expected to exist in O/Ne cores. 

\begin{figure*}
    \centering
    \includegraphics[width=1.5\columnwidth]{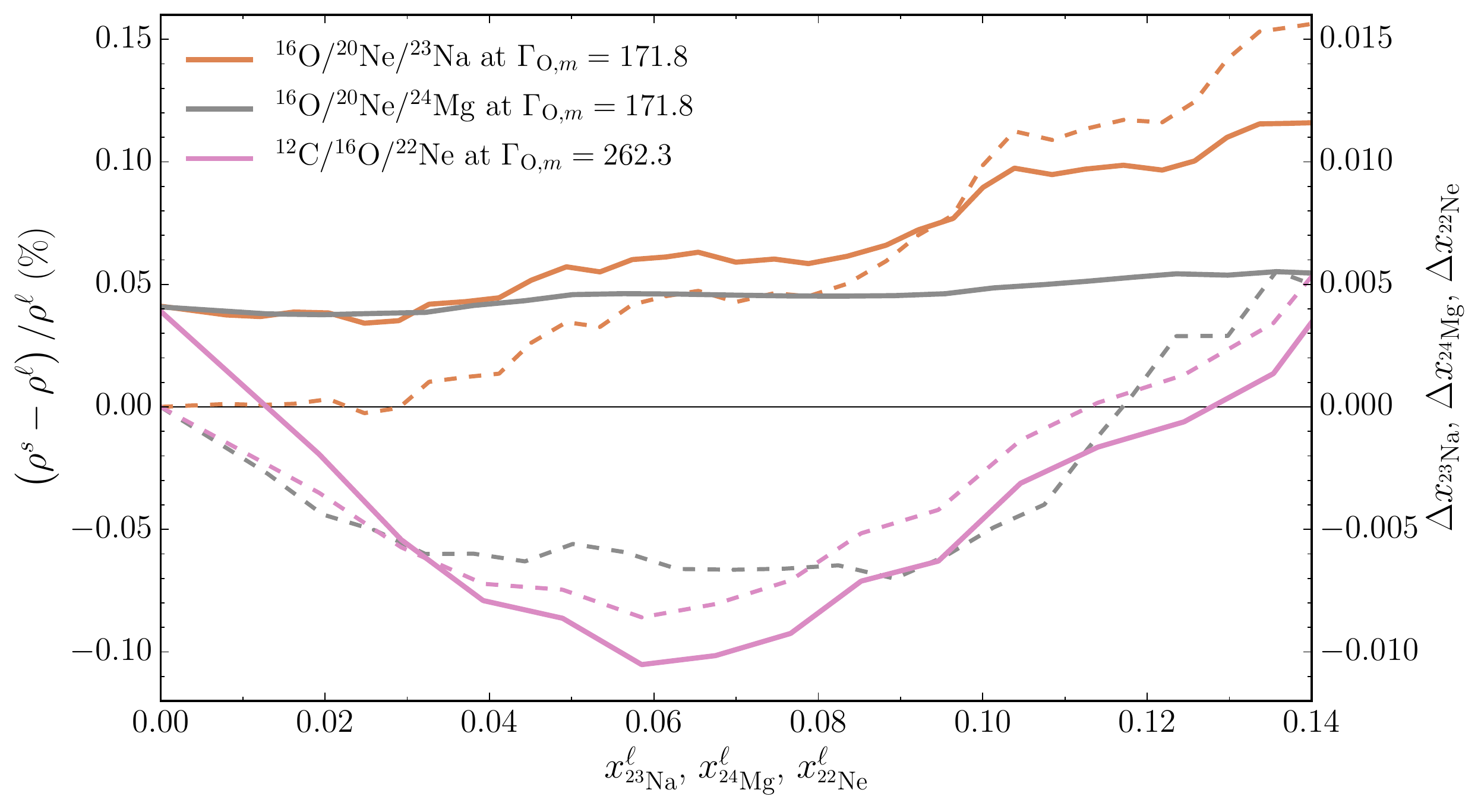}
    \caption{{\it Solid lines}: The two upper solid curves (in orange and gray) show the relative mass density difference between the liquid and solid phases at the phase transition as a function of the $^{23}$Na and $^{24}$Mg concentrations in the liquid phase for an otherwise pure O/Ne mixture. The lower solid curve (in pink) shows the same quantity as a function of the $^{22}$Ne concentration in a C/O mixture. Each curve corresponds to an integration of the relevant three-component phase diagram at constant temperature (see legend). {\it Dashed lines}: For the same mixtures and physical conditions as the solid lines, the dashed lines show the number concentration change $\Delta x = x^s - x^{\ell}$ upon solidification of the plasma. The temperature chosen for the O/Ne mixtures corresponds to $x_{\rm Ne}^{\ell}=0.35$ when no $^{23}$Na or $^{24}$Mg impurities are present; for the C/O mixture, it corresponds to $x_{\rm O}^{\ell}=0.51$ when there is no $^{22}$Ne. The individual integration points were omitted for clarity; only the moving averages are shown.}
  \label{fig:rho_comp}
\end{figure*}

That being said, it is instructive to compare the impact of $^{23}$Na ($A/Z=2.09$) on $\rho^s - \rho^{\ell}$ to that of $^{24}$Mg ($A/Z=2$). We have computed constant-temperature coexistence lines for O/Ne/Mg mixtures using the same temperatures as for our investigation of the O/Ne/Na phase diagram. In Figure~\ref{fig:rho_comp}, we show how the relative mass density difference between the coexisting liquid and solid phases changes as a function of the concentration of $^{23}$Na and $^{24}$Mg impurities in the liquid phase. In the case of $^{23}$Na, the density change (orange solid line) closely follows the $^{23}$Na concentration change (orange dashed line). The more the solid is enriched in $^{23}$Na, the heavier it is compared to the liquid. In the case of $^{24}$Mg, the density difference (gray solid line) is virtually independent of the $^{24}$Mg concentration even though substantial $^{24}$Mg concentration changes are predicted (grey dashed line). As explained above, this is because $^{24}$Mg provides the same mass per electron as $^{16}$O and $^{20}$Ne.

Figure~\ref{fig:rho_comp} also shows the case of $^{22}$Ne in a C/O mixture \citep{blouin2021a}. The concentration changes of $^{22}$Ne in C/O (dashed pink line) and of $^{24}$Mg in O/Ne (dashed gray line) are nearly identical. This can be explained by the strong similarity between the charge ratios of C/O/Ne and O/Ne/Mg mixtures (6/8/10 and 8/10/12). However, their impacts on $\rho^s - \rho^{\ell}$ are diametrically different: $^{22}$Ne's two extra neutrons imply a strong sensitivity of $\rho^s - \rho^{\ell}$ on the $^{22}$Ne concentration change. Note that the amplitude of the variations in $\rho^s - \rho^{\ell}$ (solid lines) for a given solid--liquid concentration change (dashed lines) are about twice as high for $^{22}$Ne as for $^{23}$Na, which simply reflects the fact that $^{23}$Na only has one extra neutron while $^{22}$Ne has two.

\subsection{Impact of impurities on the O/Ne phase diagram}
\label{sec:impurities}
Having computed accurate O/Ne/Na and O/Ne/Mg phase diagrams, we can now investigate the impact of Na and Mg impurities on the O/Ne phase diagram. Current evolution models rely on the two-component phase diagram to model O/Ne phase separation in ultramassive white dwarfs \citep{camisassa2019}, and it remains unclear to which extent this is a valid approximation. \citet[Table~4]{camisassa2019} present a limited investigation of this question using the semi-analytic model of \cite{medin2010}. They quantified how trace species affect O/Ne phase separation for one specific composition. Here, we use our O/Ne/Na and O/Ne/Mg phase diagrams to look at a broader range of compositions and to also study the effect of impurities on the crystallization temperature.

\begin{figure*}
    \centering
    \includegraphics[width=\columnwidth]{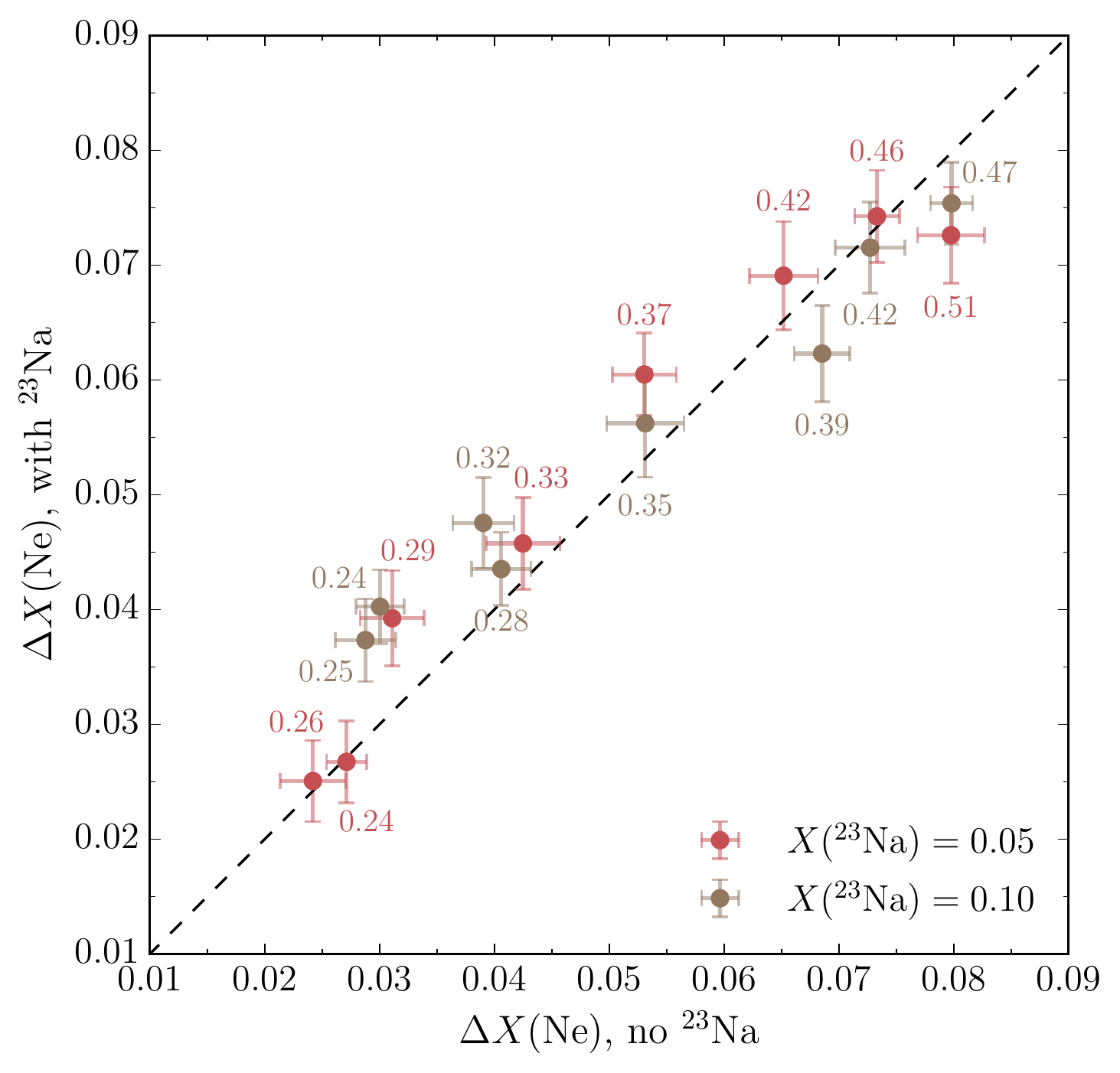}
    \includegraphics[width=\columnwidth]{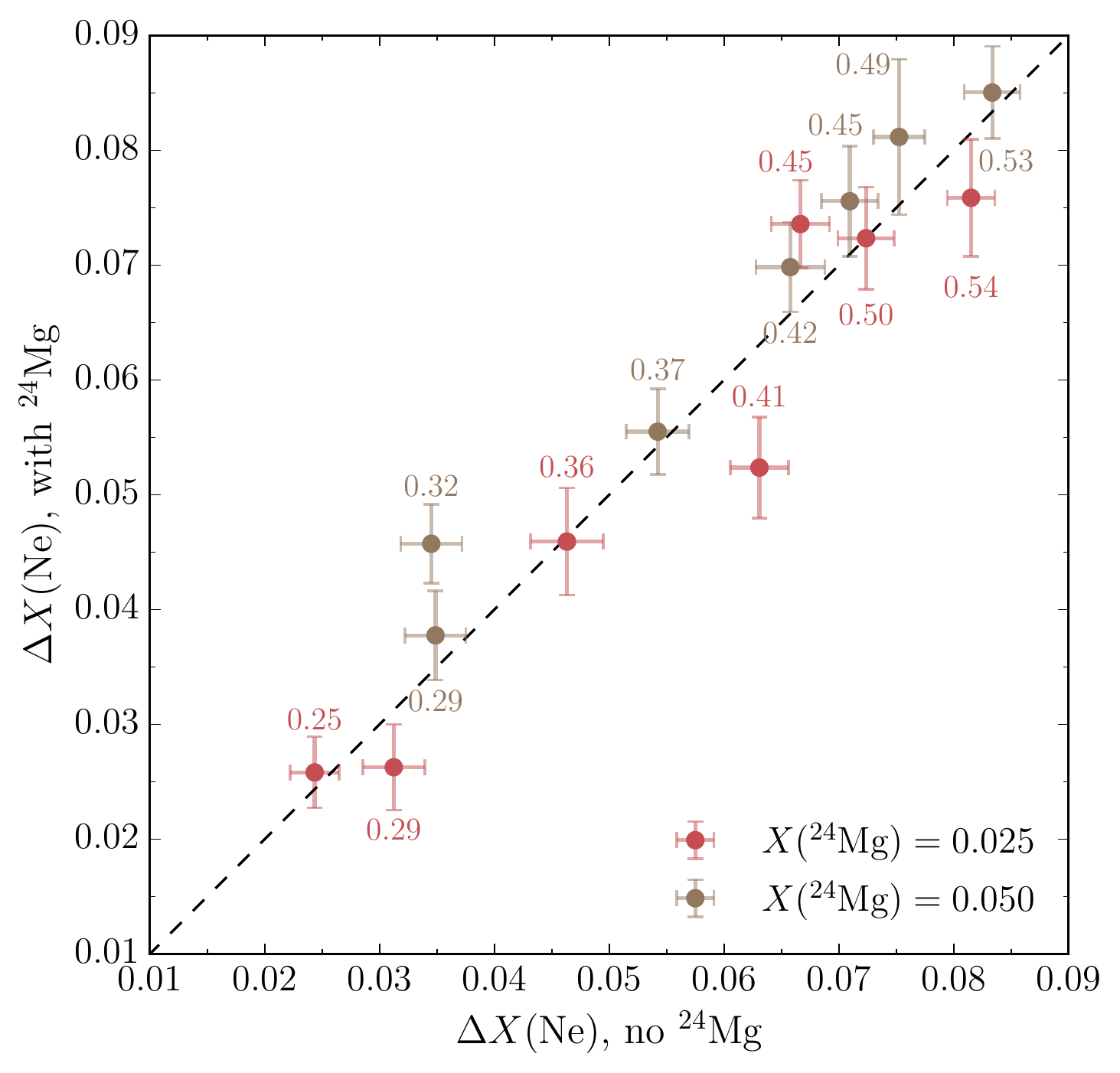}
    \caption{Effect of Na and Mg impurities on O/Ne phase separation. The horizontal axis gives the composition change upon solidification for a pure O/Ne mixture (as given by Figure~\ref{fig:ONe}). The vertical axis gives the composition change if Na (left panel) or Mg (right panel) impurities are included. Each point corresponds to a fixed Ne mass fraction in the liquid phase, the value of which is specified by a label next to each point. We show the impact of two different impurity mass fractions for both Na and Mg. Note that unlike previous figures, here we use mass fractions instead of number concentrations.}
  \label{fig:dx_perturb}
\end{figure*}

\begin{figure*}
    \centering
    \includegraphics[width=\columnwidth]{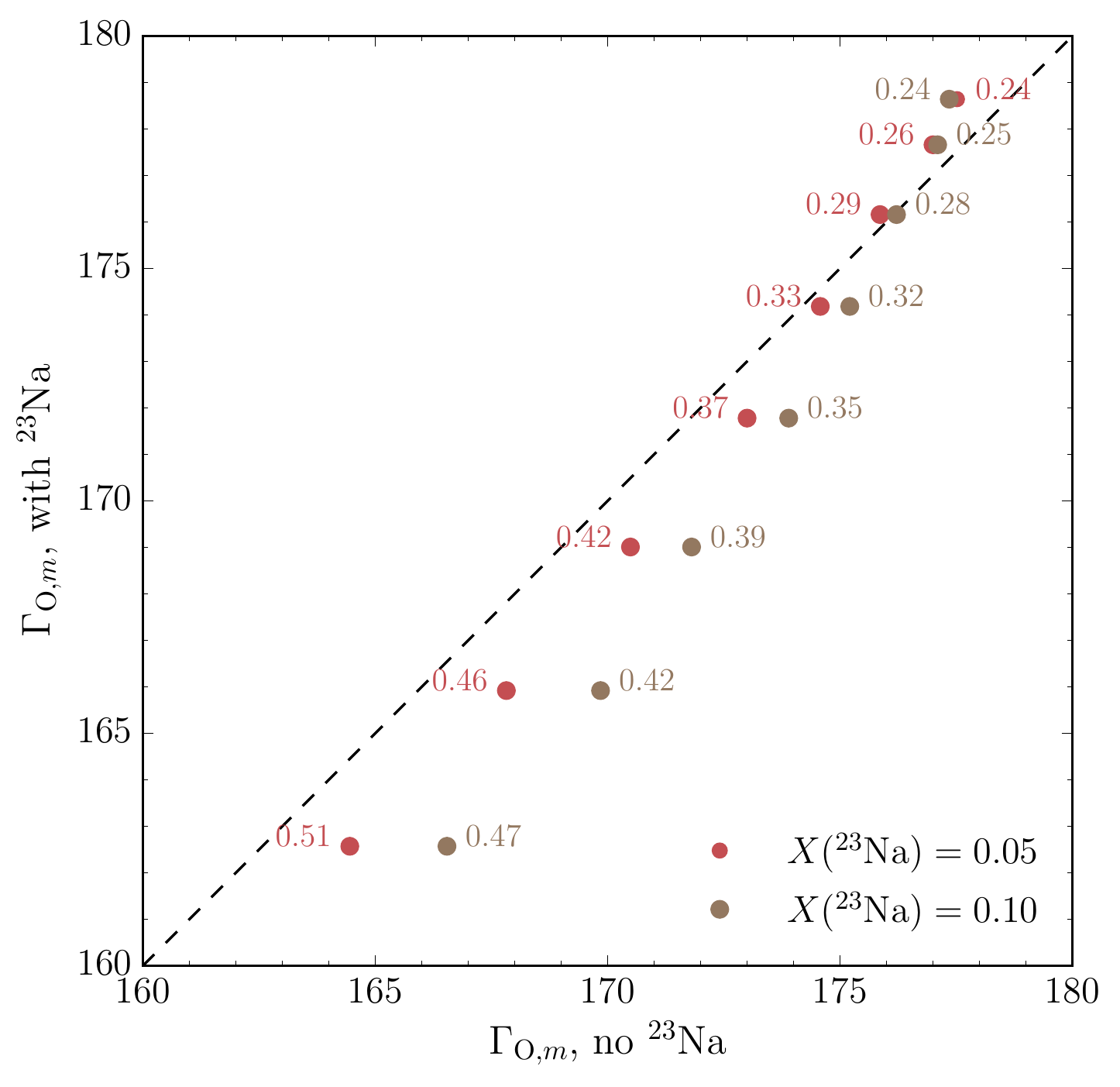}
    \includegraphics[width=\columnwidth]{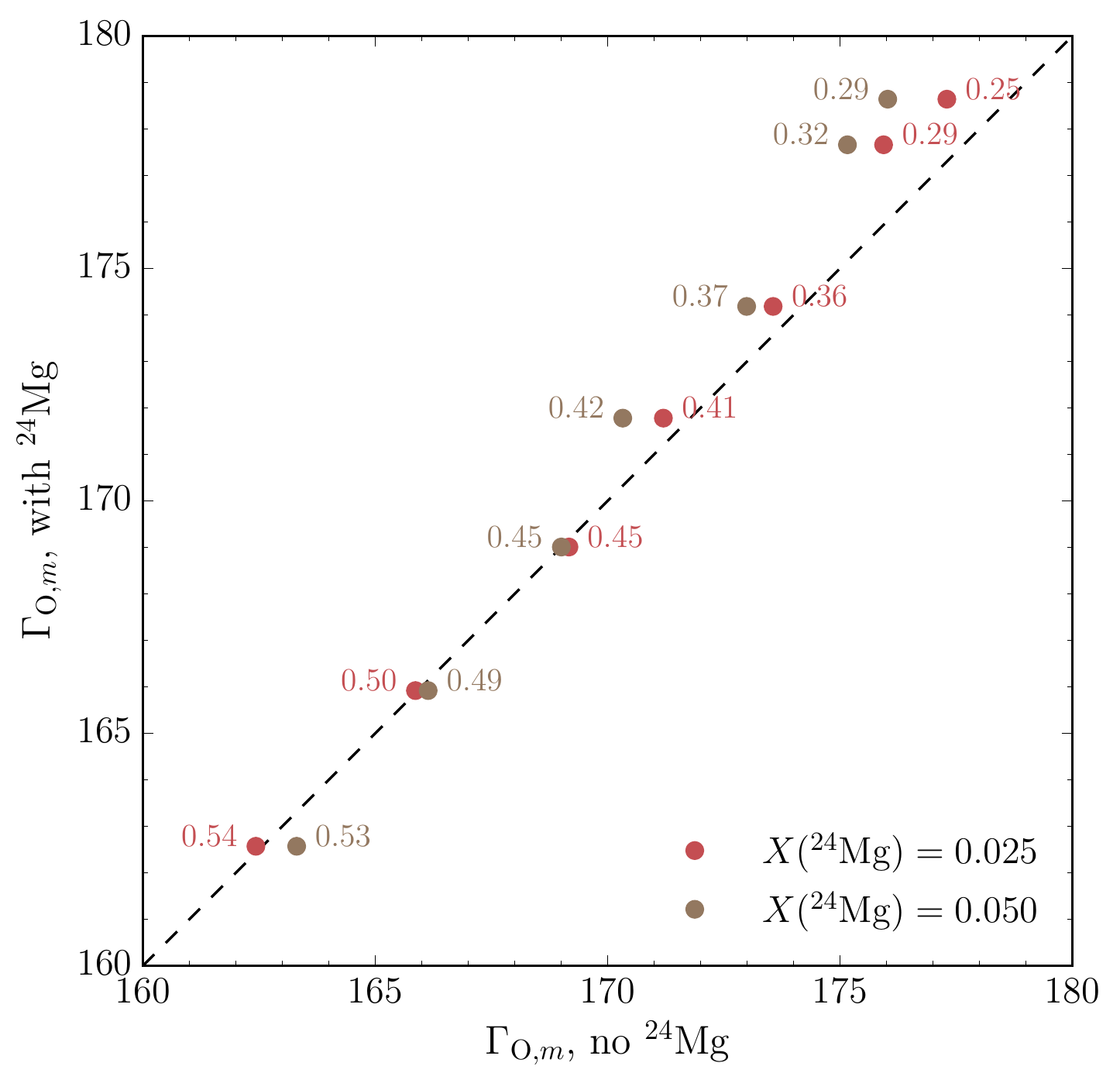}
    \caption{Effect of Na and Mg impurities on the crystallization temperature (expressed as $\Gamma_{\rm O}$) of an O/Ne mixture. The horizontal axis gives the crystallization temperature of a pure O/Ne mixture, and the vertical axis gives the crystallization temperature if Na (left panel) or Mg (right panel) impurities are included. Each point corresponds to a fixed Ne mass fraction in the liquid phase, the value of which is specified by a label next to each point. We show the impact of two different impurity mass fractions for both Na and Mg. }
  \label{fig:gamma_perturb}
\end{figure*}

Figure~\ref{fig:dx_perturb} illustrates how Na (left panel) and Mg (right panel) impurities affect O/Ne phase separation. For a set of $X({\rm Ne})$ mass fractions in the liquid phase (identified by labels next to each point), we compare the solid--liquid composition change $\Delta X ({\rm Ne})$ between the case where no impurities are included (horizontal axis) and the case where they are (vertical axis).\footnote{We use mass fractions in Figure~\ref{fig:dx_perturb} and~\ref{fig:gamma_perturb} (instead of number concentrations) to facilitate the comparison with stellar evolution codes, where this is the usual way of reporting abundances. Conversely, phase diagrams are usually reported in terms of number concentrations.} If the binary O/Ne phase diagram were a perfect representation of O/Ne phase separation in O/Ne/Na and O/Ne/Mg mixtures, all points would line up on the dashed lines.

Figure~\ref{fig:dx_perturb} shows that, within the statistical uncertainties of our calculations, all points are consistent with the dashed lines. We can therefore conclude that O/Ne phase separation is not markedly altered by Na or Mg impurities in O/Ne cores, and that the binary O/Ne phase diagram can safely be used to implement this physical process in white dwarf models. Ideally, a four-component O/Ne/Na/Mg phase diagram would be needed to assess the combined effect of Na and Mg impurities. However, Figure~\ref{fig:dx_perturb} reveals no important deviation from the two-component case even at Na and Mg abundances that significantly exceed their expected values [\cite{schwab2021b} give $X(^{23} {\rm Na})=0.060$ and $X(^{24} {\rm Mg})=0.026$]. This strongly supports using the two-component O/Ne phase diagram to model O/Ne phase separation in white dwarf codes.

Figure~\ref{fig:gamma_perturb} is similar to Figure~\ref{fig:dx_perturb}, but instead of measuring the effect of impurities on $\Delta X ({\rm Ne})$, we now look at their impact on the crystallization temperature. For a set of $X({\rm Ne})$ mass fractions in the liquid phase (identified by labels next to each point), we compare the crystallization temperature between the case where no impurities are included (horizontal axis) and the case where they are (vertical axis), both for Na (left panel) and Mg (right panel) impurities. 

Deviations from the two-component results are apparent; the points depart from the dashed lines.\footnote{Contrary to Figure~\ref{fig:dx_perturb}, we could not include error bars since temperatures are not direct outputs of our Monte Carlo simulations (unlike abundances) and are instead given by the integration of the Clapeyron equation. Nevertheless, given their systematic nature, the departures from the dashed lines are almost certainly real.} In the case of Na, the qualitative behavior of those deviations can be understood from an inspection of Figure~\ref{fig:ONeNa}. For example, Figure~\ref{fig:ONeNa} shows that both a $(x_{\rm O},x_{\rm Ne},x_{\rm Na}) \approx (0.75,0.25,0.00)$ and a $(x_{\rm O},x_{\rm Ne},x_{\rm Na}) \approx (0.65,0.25,0.10)$ mixture lie very close to the liquidus at $\Gamma_{{\rm O}} = 176.2$. This is consistent with Figure~\ref{fig:gamma_perturb}, which shows that the crystallization temperature is virtually unchanged by the addition of 10\% of Na at $X({\rm Ne}) \approx x_{\rm Ne} \approx 0.25$. At higher Ne concentrations (lower $\Gamma_{{\rm O},m}$), the shape of the O/Ne/Na phase diagram changes and increasing $x_{\rm Na}$ while keeping $x_{\rm Ne}$ fixed means ending up above the liquidus in Figure~\ref{fig:ONeNa}. This explains why the two-component phase diagram systematically overestimates $\Gamma_{{\rm O},m}$ at high Ne abundances (Figure~\ref{fig:gamma_perturb}). 

Fortunately, the Ne abundances where departures from the binary phase diagram predictions are the strongest (Figure~\ref{fig:gamma_perturb}) are substantially above the expected $X({\rm Ne}) \approx 0.3$ mass ratio of O/Ne cores \citep{camisassa2019,schwab2021b}. We can therefore conclude that using the two-component O/Ne phase diagram to predict the crystallization temperature of the real multicomponent mixture in O/Ne cores should not result in an error of more than $1-2$\% on $\Gamma_{{\rm O},m}$, which is negligible compared to other sources of uncertainties in white dwarf cooling models.

\section{Conclusions}
\label{sec:conclu}
We have calculated the most reliable version to date of the O/Ne phase diagram and provided analytic fits to facilitate its implementation in ultramassive white dwarf models, where O/Ne phase separation needs to be taken into account. We have explicitly shown that this binary phase diagram is a very good approximation of the phase diagram of the more complex multicomponent ionic mixture that characterizes ultramassive O/Ne white dwarfs, which also includes significant amounts of Na and Mg.

Our detailed investigation of the O/Ne/Na phase diagram has revealed that a $^{23}$Na distillation process analogous to $^{22}$Ne distillation in C/O white dwarfs cannot take place in ultramassive O/Ne white dwarfs (at least unless our picture of the core composition of those objects changes considerably). Without this efficient $^{23}$Na transport mechanism, we conclude that no known process can transport enough $^{23}$Na to the central layers of ultramassive white dwarfs to trigger an electron capture white dwarf collapse as proposed by \cite{caiazzo2021}.

\acknowledgements
We thank Didier Saumon for helpful comments that have improved the clarity of this manuscript and the anonymous referee for a rapid report.

Research presented in this article was supported by the Laboratory Directed Research and Development program of Los Alamos National Laboratory under project number 20190624PRD2. This work was performed under the auspices of the U.S. Department of Energy under Contract No. 89233218CNA000001.

\bibliography{references}{}
\bibliographystyle{aasjournal}

\end{document}